\documentclass[10pt, paper=a4, UKenglish]{article}
\usepackage{graphicx}
\def\Title#1{\begin{center} {\Large #1 } \end{center}}
\def\Author#1{\begin{center}{ \sc #1} \end{center}}
\def\Address#1{\begin{center}{ \it #1} \end{center}}

\newcommand\pubblock{\rightline{\begin{tabular}{l} Proceedings of the CTD/WIT 2019\\ \pubnumber\\
         \pubdate  \end{tabular}}}

\newenvironment{Abstract}{\begin{quotation} \begin{center} 
             \large ABSTRACT \end{center}\bigskip 
      \begin{center}\begin{large}}{\end{large}\end{center} \end{quotation}}

\newenvironment{Presented}{\begin{quotation} \begin{center} 
             PRESENTED AT\end{center}\bigskip 
      \begin{center}\begin{large}}{\end{large}\end{center} \end{quotation}}

\def\Acknowledgements{\bigskip  \bigskip \begin{center} \begin{large}
      \bf ACKNOWLEDGEMENTS \end{large}\end{center}}

\usepackage{color}
\usepackage{lineno}
\usepackage{subfig}
\usepackage{hyperref}

\usepackage{caption}

\newcommand\pubnumber{PROC-CTD19-091}

\newcommand\pubdate{\today}

\def \affiliation{
On behalf of the PANDA Collaboration, \\
Institut f{\"u}r Kernphysik (IKP) \\
Forschungszentrum J{\"u}lich, J{\"u}lich}

\newcommand{\conference}{Connecting the Dots and Workshop on Intelligent Trackers (CTD/WIT 2019)\\
Instituto de F\'isica Corpuscular (IFIC), Valencia, Spain\\ 
April 2-5, 2019}

\usepackage{fancyhdr}
\usepackage{ragged2e}
\usepackage{gensymb}
\definecolor{mygrey}{RGB}{105,105,105}
\begin{document}


\large
\begin{titlepage}
\pubblock

\vfill
\Title{Machine Learning for Track Finding at PANDA}
\vfill

\Author{W.Esmail, T.Stockmanns, and J.Ritman}
\Address{\affiliation}
\vfill
\begin{Abstract}
\begin{flushleft}
\justify
We apply deep learning methods as a track finding algorithm to the PANDA Forward Tracking Stations (FTS).
The problem is divided into three steps:
\newline
The first step relies on an Artificial Neural Network (ANN) that is trained as a binary classifier to build track segments in three different parts of the FTS, namely FT1,FT2, FT3,FT4, and FT5,FT6. The ANN accepts hit pairs as an input and outputs a probability that they are on the same track or not. The second step builds 3D track segments from the 2D ones and is based on the geometry of the detector. The last step is to match the track segments from the different parts of the FTS to form a full track candidate, and is based on a Recurrent Neural Network (RNN). The RNN is used also as a binary classifier that outputs the probability that the combined track segments are a true track or not. The performance of the algorithm is judged based on the purity, efficiency and the ghost ratio of the reconstructed tracks. The purity specifies which fraction of hits in one track come from the correct particle. The correct particle is the particle, which produces the majority of hits in the track. The efficiency is defined as the ratio of the number of correctly reconstructed tracks to all generated tracks.
\end{flushleft}

\end{Abstract}

\vfill

\begin{Presented}
\conference
\end{Presented}
\vfill
\end{titlepage}
\def\thefootnote{\fnsymbol{footnote}}
\setcounter{footnote}{0}
%

\normalsize 


\section{Introduction}
\label{intro}

The PANDA detector (anti\textbf{P}roton \textbf{AN}nihilation at \textbf{DA}rmstadt) is one of the key experiments at FAIR (\textbf{F}acility for \textbf{A}ntiproton and \textbf{I}on \textbf{R}esearch) in Darmstadt, Germany \cite{panda}. The PANDA Collaboration will address various questions related to the strong interaction. Anti-protons with momenta from 1.5 to 15 GeV/c will collide with a fixed proton target. The anti-proton proton interaction allows to study a wide variety of physics questions as particles of all quantum numbers can be produced. The research targets of the experiment will include \cite{physics}:
\begin{itemize}
  \item Hadron Spectroscopy: The excitation spectrum of a large number of baryons is not fully
understood; the experimental measurements of some states do not agree very well with the theoretical predictions of the quark model, while other states are not seen at all.
  \item Gluonic Excitations: The Standard Model allows bound states with excitated gluons – either as gluonic excitations of standard hadrons (hybrids) or gluon states (glueballs) without any valence quarks. None of the predicted gluonic states have yet been seen clearly and unambiguously in experiments.
  \item Nucleon Structure: Different electromagnetic processes at PANDA will allow to study nucleon structure observables, this can be achieved by Drell-Yan processes ($p\bar{p} \rightarrow e^{+}e^{-}X $) that give access to the time-like region of the proton form factors.
  \item Hypernuclei: The experiment will investigate the effect of hadrons implemented into nuclear matter. Hyperons can be implemented into nuclei replacing a proton or a nucleon, since they decay weakly they form quasi-stable hypernuclei which can be studied at PANDA. 
\end{itemize}

\begin{figure}[ht]
\centering
     \includegraphics[width=0.9\textwidth]{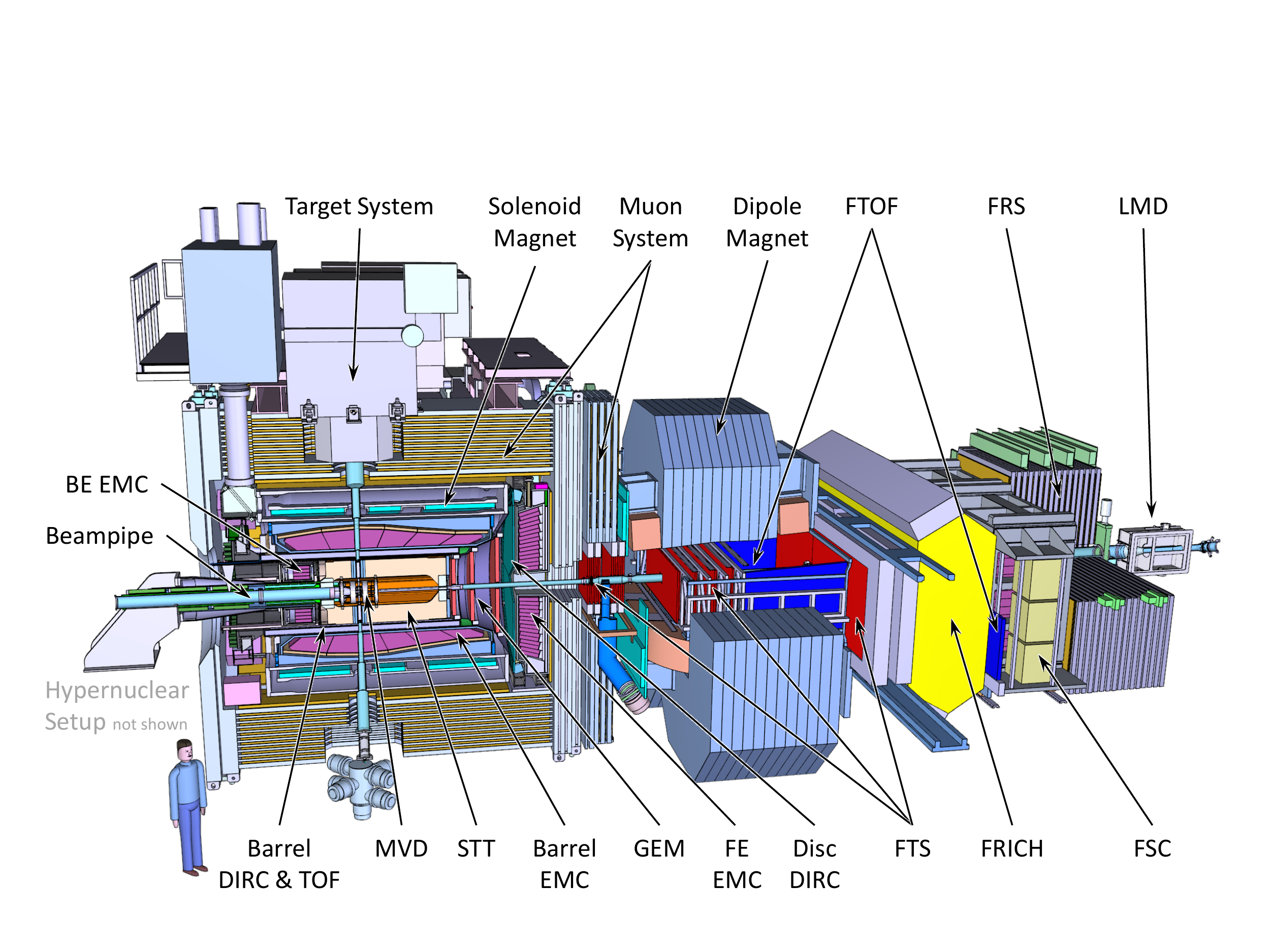}
      \caption{The PANDA full setup.}
       \label{fig:panda_setup}
\end{figure}

The rich physics program of PANDA requires a multi-purpose detector system which is illustrated in figure\ref{fig:panda_setup}. The PANDA sub detector system is arranged into two parts, the target spectrometer which covers the central region around the interaction point and instruments nearly the full 4$\pi$ solid angle, and the forward spectrometer downstream of the interaction to measure forward boosted particles at low polar angles. The forward spectrometer aims to measures particles at small polar angles $\theta$ below 5$^{\circ}$ in the vertical and 10$^{\circ}$ in the horizontal plane. Its magnetic field with a maximum bending power of 2 Tm is provided by a dipole magnet. For the measurement of particle momenta based on the deflection of their trajectories in the magnetic field, the \textbf{F}orward  \textbf{T}racking  \textbf{S}ystem (FTS) is foreseen \cite{fts}.
\newline

\begin{figure}[!htb]
  \centering
  \subfloat[]{\includegraphics[width=0.45\linewidth]{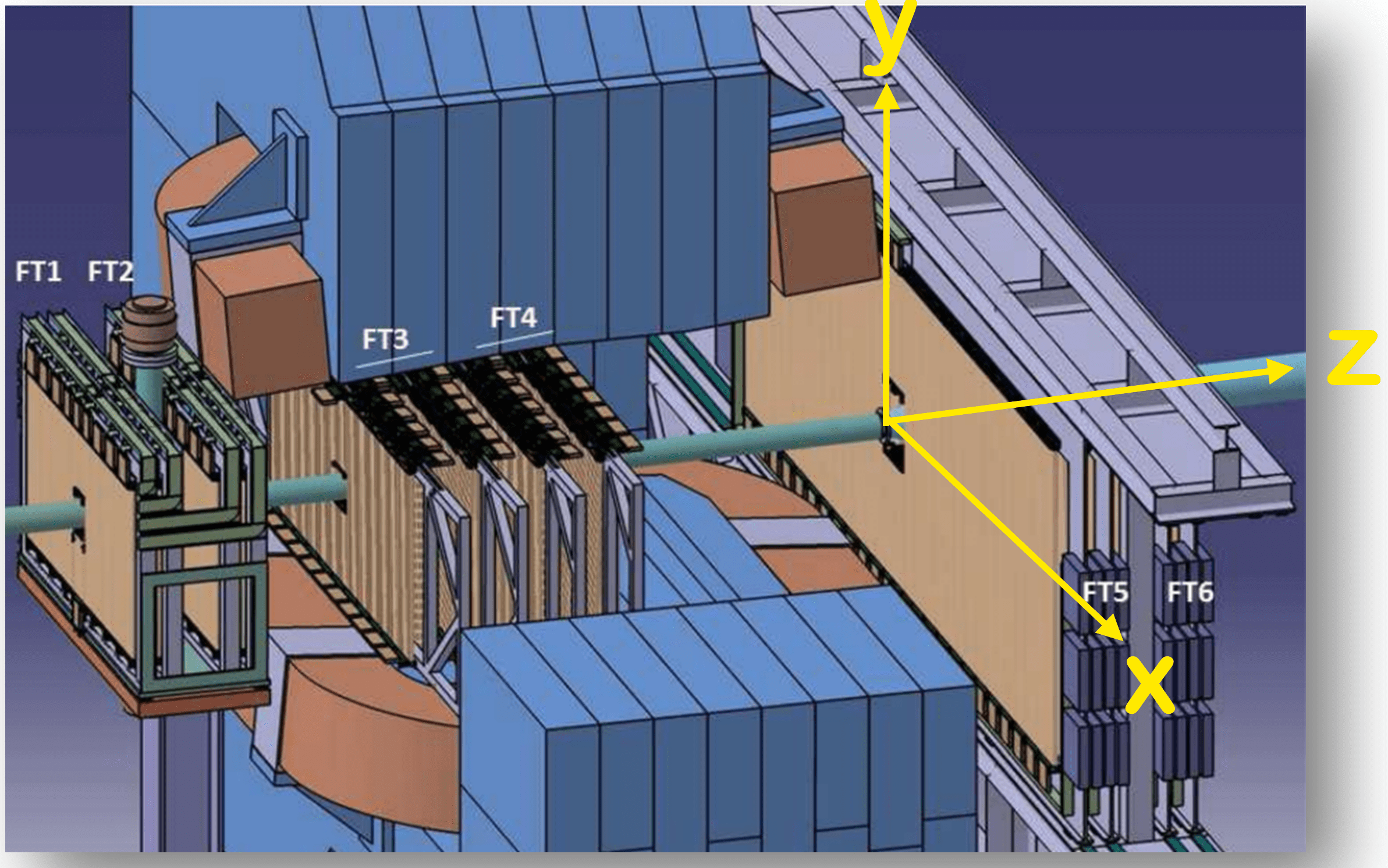}}
  \qquad
  \subfloat[]{}{\includegraphics[width=0.40\linewidth]{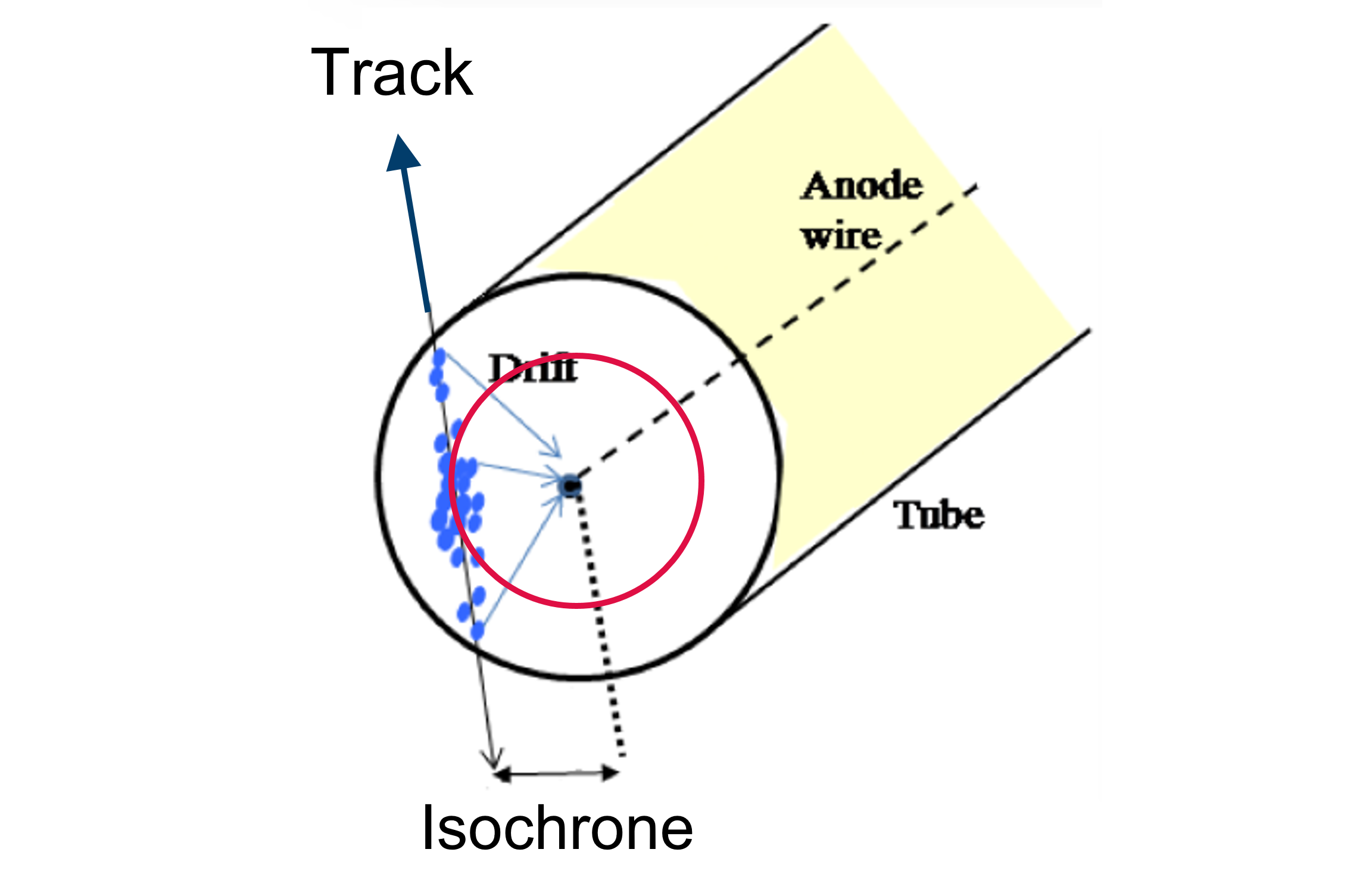}}
  \caption{(a) Schematic view of the PANDA Forward Tracking System \cite{fts} (b) Schematic view of the straw tube and the operation of particle tracking, the red circle define the isochrone radius where the particle track is tangent to it \cite{straw}.}
  \label{fig:fts_drift}
\end{figure}

As shown in figure \ref{fig:fts_drift} (a), six FTS tracking stations are planned. The first two, dubbed FT1 and FT2, are located in front of the magnet, the last two, FT5 and FT6, are located after the magnet, while FT3 and FT4 are located inside the dipole magnet. Each tracking station is composed of four double layers of straw tubes. The first and last double layer are aligned vertically, while the second and the third double layer are tilted by +5$^{\circ}$ and -5$^{\circ}$ with respect to the vertical direction, respectively.

Straw tubes are gaseous drift detectors usually operating in the proportional region. They consist of a gas filled conducting tube and an anode wire at high voltage stretched along the tube axis. When charged particles pass through the tube, they ionize the gas creating electron-ion pairs. Charged particles that traverse the straw tubes are tracked by measuring the drift time of the ionization electrons to the anode wire. Using the drift time information, the distance of closest approach of the charged particle track to the anode wire is determined. This distance is the main information for charged particle tracking using straw tube detectors. This distance is also called the ioschrone radius, figure \ref{fig:fts_drift} (b).

The aim of this study is to assess the ability of deep learning techniques to reconstruct charged particle tracks in the FTS.

\section{Tracking Model}
\begin{figure}[ht]
\centering
     \includegraphics[width=0.7\textwidth]{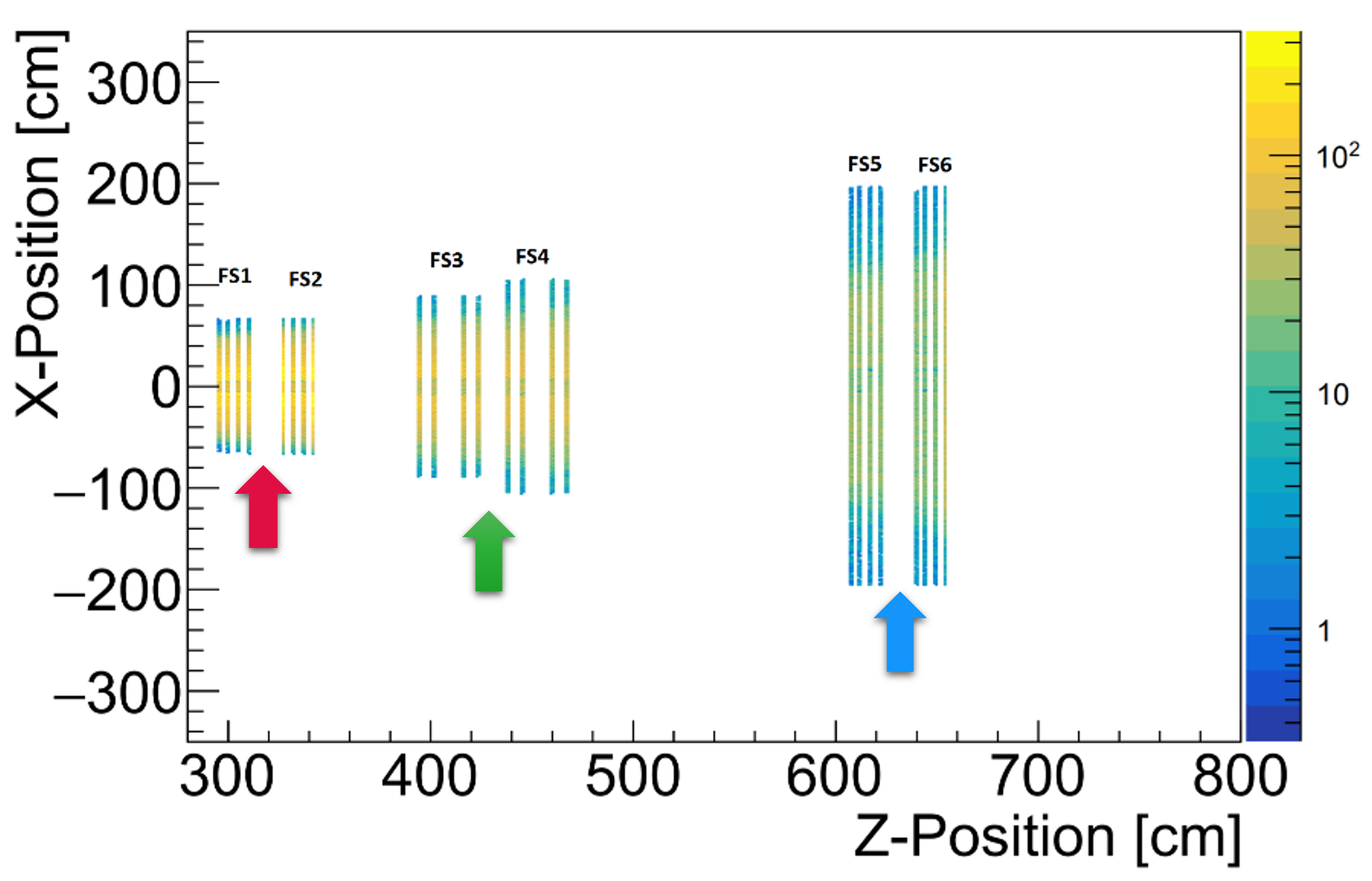}
      \caption{X-Z projection of detector hits.}
       \label{fig:ftsxz}
\end{figure}

In this section we describe our current approach to apply deep learning techniques to the problem of track finding in the FTS. We break the problem into three sub-problems as follows:
\begin{enumerate}
    \item \textbf{Creating Track Segments}: Apply a feed forward \textbf{N}eural \textbf{N}etwork (NN) to create 2D track segments (seeds) in different stations, specifically in \textcolor{red}{(FT1+FT2)}, \textcolor{green}{(FT3+FT4)}, and \textcolor{blue}{(FT5+FT6)} as illustrated in figure \ref{fig:ftsxz}.
    \item \textbf{Addition of Skewed Layers}: Extend 2D track segments into 3D by using the skewed layers.
    \item \textbf{Connecting Track Segments}: Match the track segments to a full track candidate using a \textbf{R}ecurrent \textbf{N}eural \textbf{N}etwork (RNN).
\end{enumerate}

For the purpose of preparing a training dataset for the proposed algorithm, the BOX
\textbf{M}onte-\textbf{C}arlo (MC) event generator of PandaRoot \cite{pandaroot} is used to simulate tracks of 5 muons per event emitted isotropically from the target at polar angles ranging from 0.5$^{\circ}$ to 10$^{\circ}$, with a momentum range from 1 to 10 GeV/c. The result of the simulation is a collection of detector hits that represent the wire position (i.e the center of the straw tube), and the ioschrone radii associated to each hit.

\subsection{Creating track segments}
The first step of the algorithm is to combine hits in \textcolor{red}{(FT1+FT2)}, \textcolor{green}{(FT3+FT4)}, and \textcolor{blue}{(FT5+FT6)} to create track segments. The output of this step is a collection of 2D track segments in the three parts of the FTS. For this purpose an artificial NN implemented within the Keras framework is used \cite{keras}. The NN is trained to predict if hit pairs are on the same track or not.
The training dataset is divided into two sets, one set contains hits outside the magnetic field and the other contains hits inside the field volume.
\hfill \break
For each event, all possible combinations of hit pairs in adjacent detector layers are formed and fed to the network, the input variables to the network are:
\begin{itemize}
    \item X and Z coordinates of the hit pair for the vertical layers.
    \item Isochrones associated to each hit.
    \item The distance between hits.
\end{itemize}

The network is used as a binary classifier with a single output neuron with a sigmoid activation function. This single neuron outputs the probability that the input sample belongs to the positive class (hit pairs are on the same track). If it is true, then the output value should be close to 1, otherwise it will be tagged as a member of the negative class (hit pairs belonging to different tracks).
\hfill\break
To train our classifier the binary cross-entropy error is minimized \cite{entropy}. The binary cross-entropy formula is defined as follows:

\begin{equation}
    L(y,\hat{y}) = -y\ log\hat{y} - (1-y)\ log(1-\hat{y})
\end{equation}

where y is the label (truth value), and $\hat{y}$ is the network prediction.
\hfill\break

We found in our study that the best results of validation accuracy was obtained by the combination of five fully connected layers one after another. In addition to that, we applied a 50\% dropout \cite{dropout} to each layer to prevent our model from over-fitting. Two NN are used for this step, one is trained for stations outside the dipole magnet, the other network is trained for the stations inside the dipole magnet. They score 98.5\% and 97\% accuracy, respectively after 1000 iterations of training.

\begin{figure}[!htb]
  \centering
  \includegraphics[width=0.99\linewidth]{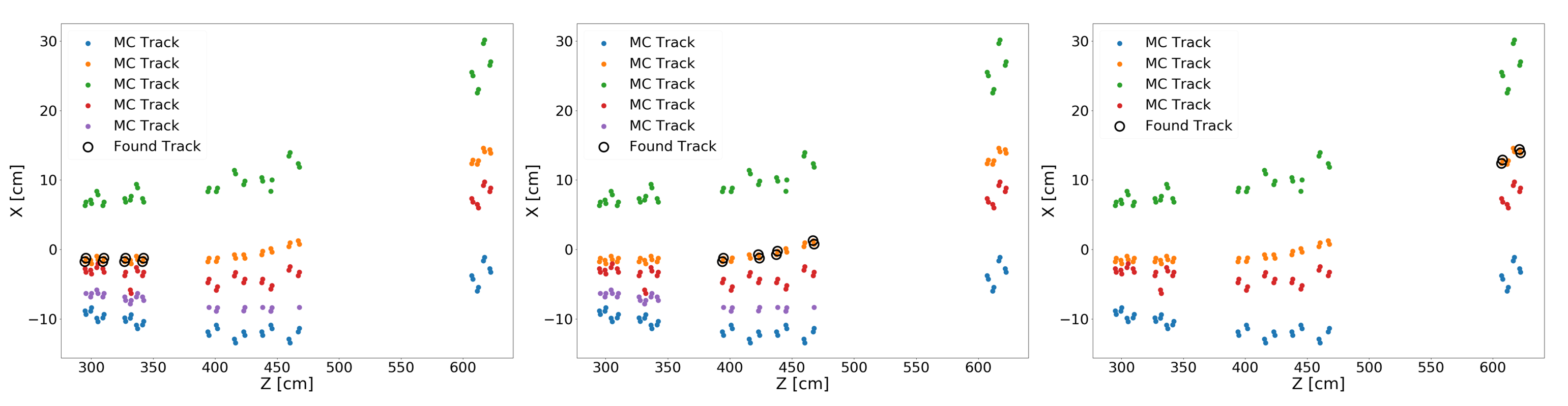}
  \caption{Pictorial representation of the network output of an event. Different colors indicate different MC tracks. The black circles indicate the output of the network for (FT1,FT2), (FT3,FT4), and (FT5,FT6) from left to right respectively.}
  \label{fig:pic_rep}
\end{figure}

\hfill \break
The output probability of the network can be used to connect hits together in a simple way to form track segments. In other words, if the probability($hit_{1}, hit_{2}$) passes the probability cut, and probability($hit_{2}, hit_{3}$) also passes the probability cut, then $hit_{1}$, $hit_{2}$, and $hit_{3}$ are all on the same track. The probability cut used for this study is 90\%. Figure \ref{fig:pic_rep} shows a pictorial representation of the network prediction.

\subsubsection{Neural network performance}
To evaluate our classifier we generated another test set of events with 5 tracks as particle multiplicity using the BOX generator. We put quality criteria to identify found tracks and judge the performance of the classifier. If the found track has greater than 4 detector hits the track is preserved for further processing, otherwise the track is rejected. Then the purity is determined, where the purity specifies which fraction of hits in one track come from the matching particle. The matching particle is the particle, which produces the majority of hits in the track. 

A common assumption is that a found track matches to a true track if 80\% of its hits come from the true track.
\hfill \break
Purity reflects the quality of the reconstruction and the contamination of the tracks. Another figure of merit for the quantity of the reconstructed tracks is the efficiency. Efficiency of track reconstruction can be defined as the ratio of the number of the pure reconstructed tracks (purity \textgreater 80\%) to all tracks present. Table \ref{tab:table1} summarize the quality of the classifier. 

\begin{table}[!htb]
  \begin{center}
    \begin{tabular}{|| l | l || l | l ||}
      \hline
      \hline
        &  FT1,FT2 & FT3,FT4 & FT5,FT6 \\
      \hline
      Purity &  99\%  &  99\% & 99\% \\
      Efficiency &  96\%  &  95\% & 96\% \\
      \hline
      \hline
    \end{tabular}
    \caption{Purity and Efficiency of the neural network.}
    \label{tab:table1}
  \end{center}
\end{table}

\label{section}

\subsection{Addition of skewed layers}

The selected X-Z projection of track candidates only contains hits from the vertical layers. The Y-Z plane track motion is extracted from the skewed layers since their local frame is obtained from a rotation of the X-Y plane around the Z direction by +5$^{\circ}$ and -5$^{\circ}$. Therefore, it is possible to add the information of the track motion in the Y-Z plane looking at the skewed layer hits which are compatible with the X-Z plane track projection. The X-Z found track segments are used as seeds for this task. The addition of skewed hits can be summarized as follows:

\subsubsection{X-Z projection fitting}
\begin{figure}[!htb]
  \centering
  \includegraphics[width=0.99\linewidth]{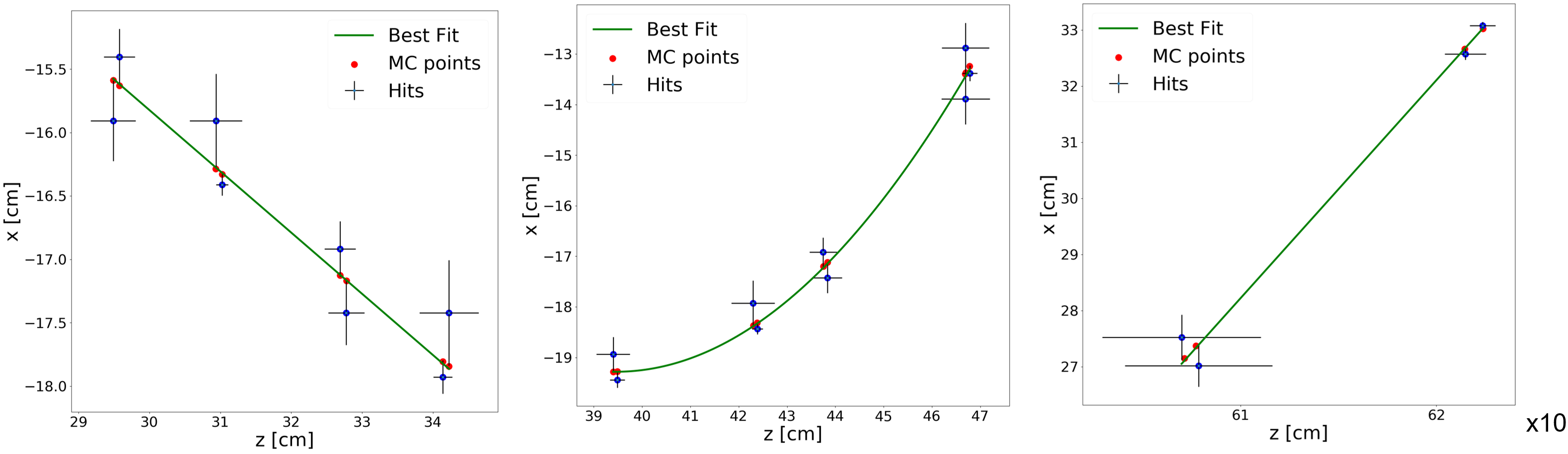}
  \caption{An example of the track segments fitting in (FT1,FT2), (FT3,FT4), and (FT5,FT6) from left to right respectively. Blue points represent detector hits, while red points represent MC points. The best fit is shown by the green curve. }
  \label{fig:pic_fit}
\end{figure}

Each X-Z projection is fitted by using an Orthogonal Distance Regression (ODR) implemented within Scipy \cite{odr} \cite{scipy}. ODR differs from Ordinary Least Squares method (OLS) by accounting for errors in observations on both the explanatory and the response variables (Z and X in this case). The errors for the two variables are assumed to be independent. 
\hfill\break
\hfill\break
Each hit is treated as a data point in X-Z coordinates, while the isochrones are treated as errors in both coordinates. A linear fitting is used for the stations outside the magnetic field, while a parabolic fit is chosen for the stations inside the magnetic field. An example of the fitting procedure is shown in figure \ref{fig:pic_fit}. The figure shows that the fit passes exactly through each MC point indicating the quality of the fit.

\subsubsection{Collect compatible hits}
\begin{figure}[!htb]
  \centering
  \includegraphics[width=0.90\linewidth]{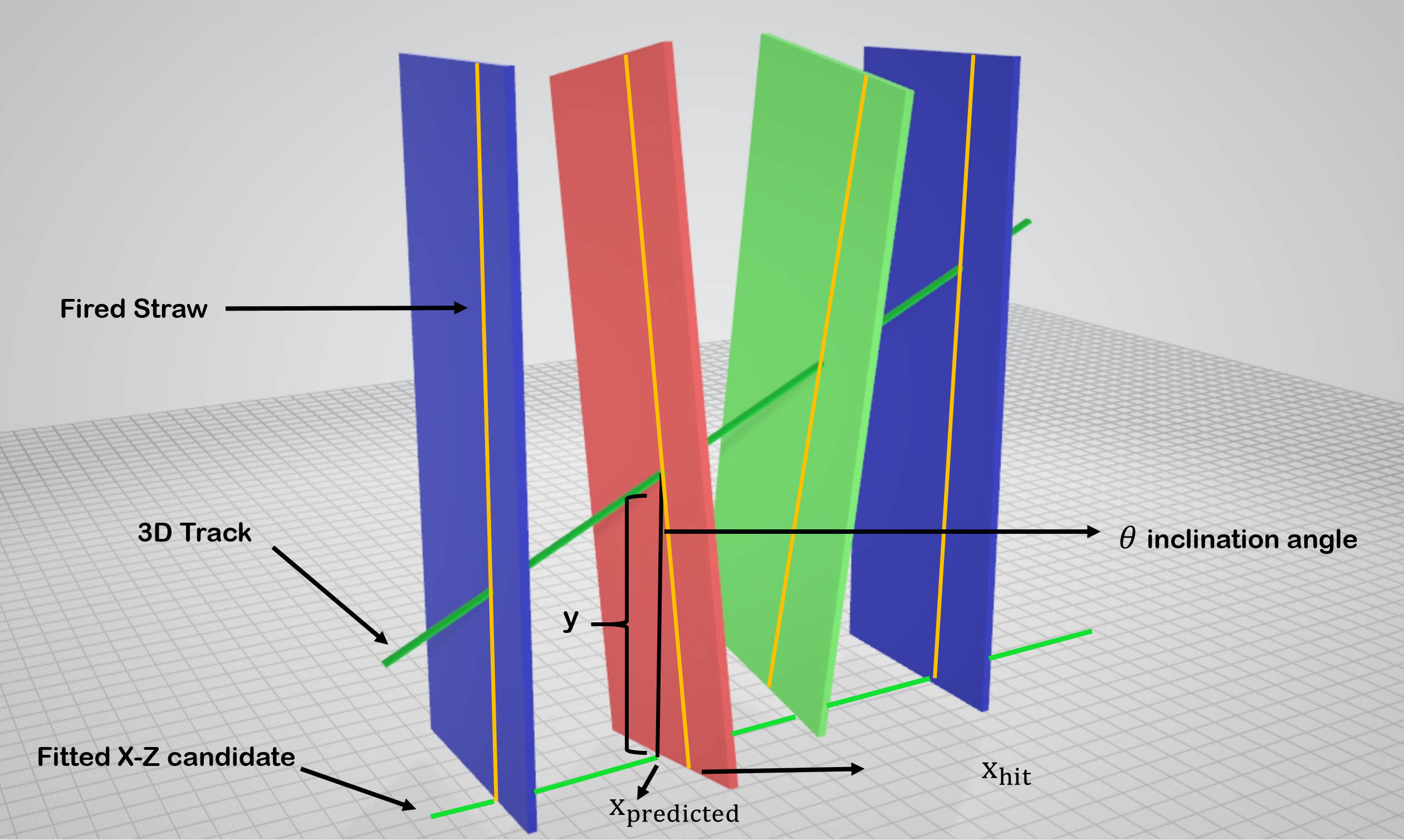}
  \caption{A drawing showing the geometrical interpretation of compatible hits based on the fitted X-Z projections. }
  \label{fig:pic_geom}
\end{figure}

For each fitted X-Z projection the predicted X position at the Z position
of skewed layers is evaluated. The distance between the skewed layer measurements and the predicted X position allows to identify a Y measurement for each skewed hit. Therefore all the hits compatible in Y with respect to the X-Z projection are collected. Due to the tilted orientation of the skewed layers the predicted X hit position in the local frame of skewed layers corresponds to:
$$
x_{predicted} = x_{hit} + y\ tan\theta
$$
where $\theta=\pm5^{\circ}$ is the angle of inclination of the skewed layers. The compatibility of the skewed hits for a given X-Z projection can be defined by looking at the slope:
$$
slope = \frac{ ( x_{predicted}-x_{hit} )\ /\ tan\theta }{z_{hit}}
$$
Therefore, a group of hits sharing the same slope value define a potential line candidate to be associated to the X-Z projection candidate. A drawing showing the detector geometry and the process of extracting a Y measurement is shown in figure \ref{fig:pic_geom}.

\begin{figure}[!htb]
  \centering
  \includegraphics[width=0.99\linewidth]{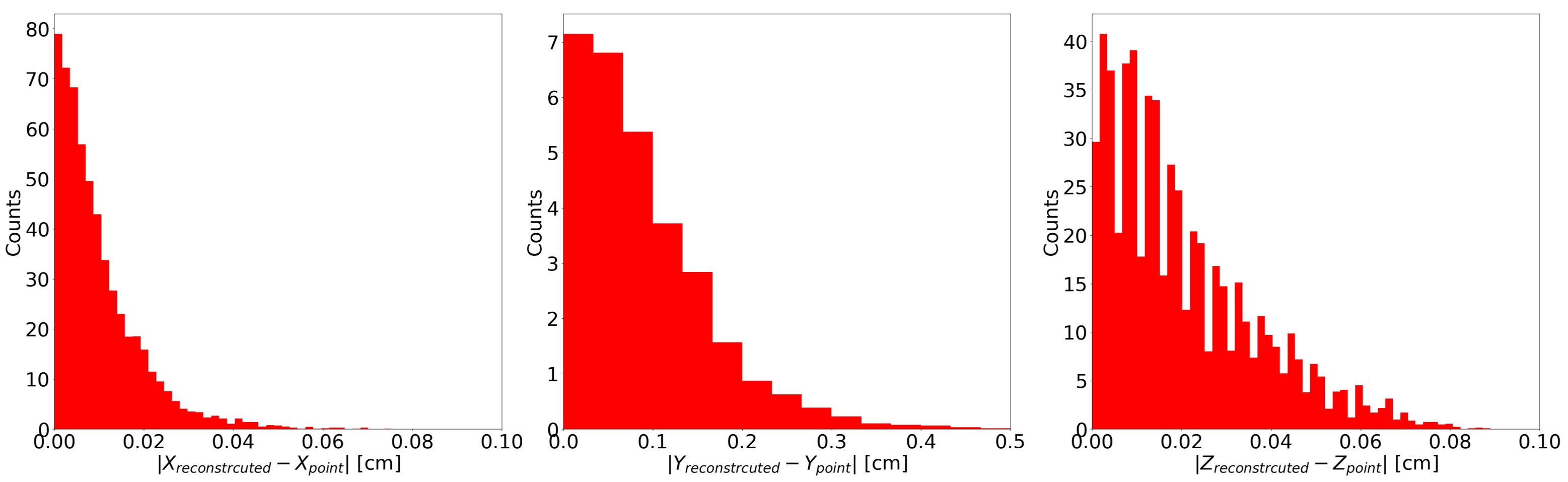}
  \caption{Distribution of the corrected hit position and the corresponding MC points for the X, Y, and Z coordinates, respectively. }
  \label{fig:pic_res}
\end{figure}

\hfill\break
\hfill\break
As mentioned above the position of the hit is always generated exactly in the middle of the straw tube (wire position). The corrected hit positions lie exactly at the points where the track is tangent to the isochrones. Since the fitted line/curve incorporates isochrones as errors on both axes, it provides the correct hit positions. This is advantageous because in the later course of the process of track reconstruction, a track fitting is made. A more accurate position of the hits used for the fitting also improves the quality of the fit. Distributions of the residua of the corrected hit position and the corresponding MC points are shown in figure \ref{fig:pic_res}. The width of the distribution in the vertical direction is roughly 0.15 cm which is comparable to the vertical spatial resolution design value ($\sigma_{y}$ $\sim$ 0.15 cm) \cite{fts}. The resolution in the horizontal resolution is 10 times better than the vertical direction.
\hfill\break
After the fitting procedure the 2D track segments extend to 3D segments. The next step is to match these segments to build a full track candidate.

\subsection{RNN Connecting the track segments}
A major characteristic of feed forward neural networks is that they have no memory. Each input shown to them is processed independently and with such networks in order to process a sequence you have to show the entire sequence to the network at once. In contrast RNNs can process sequential information, in other words they have an internal memory which captures information about what has been calculated so far. A simple block of RNN represents a loop over inputs divided into time steps, in the heart of the loop there is a feed forward neural network. The unrolled block, can be thought of as multiple copies of the same network, each passing a message to a successor. 
As particle tracks present a sequence of hits, RNN perfectly suits the task of connecting track segments. Each time step of such sequences is the specific point on the i-th layer of the FTS.

\begin{figure}
  \begin{minipage}[c]{0.40\textwidth}
    \includegraphics[width=\textwidth]{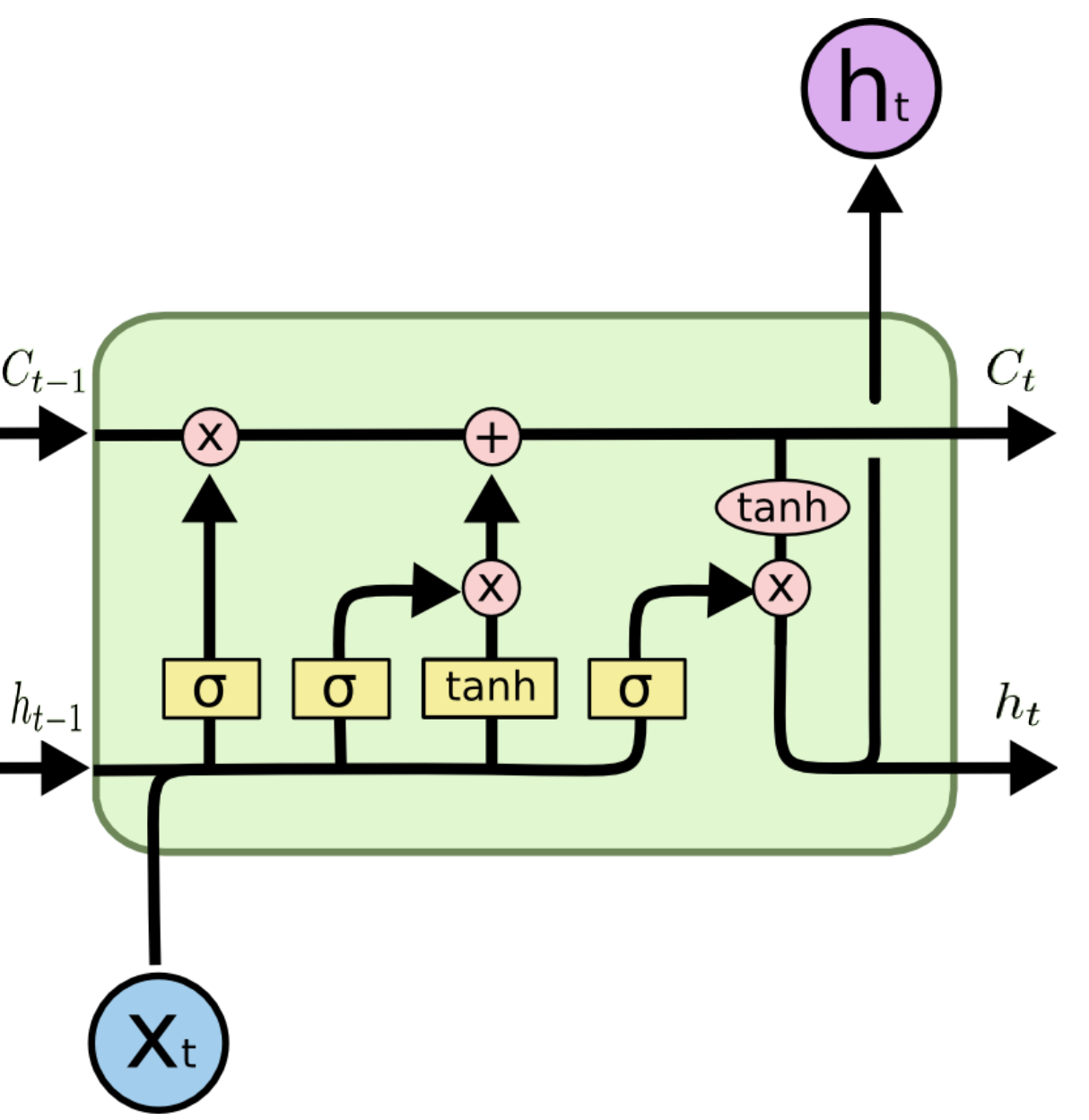}
  \end{minipage}\hfill
  \begin{minipage}[c]{0.48\textwidth}
    \caption{A LSTM cell \cite{lstmcell}. A LSTM has three of these gates, to control the cell state $C_{t}$. The "forget gate" decides what information is going to be thrown away from the cell state, this decision is made by a sigmoid $\sigma$ activation. It looks at the previous calculation $h_{t-1}$ and the input $x_{t}$ and outputs a number between 0 and 1 for each number in the cell state $C_{t-1}$. The "input gate layer" decides what new information is going to be stored in the cell state. This is controlled by a sigmoid activation that decides which values will be update, and a $tanh$ activation that creates a vector of new candidate values, that will be added to the cell state. Finally, the "output gate" decides what cell is going to the output, which will be based on the cell state. This is controlled by a sigmoid and $tanh$ activations.} 
    \label{fig:pic_lstm}
  \end{minipage}
\end{figure}

A \textbf{L}ong \textbf{S}hort-\textbf{T}erm \textbf{M}emory (LSTM) \cite{lstm} implemented within the Keras framework is used for matching track segments. A LSTM is a special kind of RNN capable of learning long-term dependencies. LSTMs also have the same chain-like structure as simple RNN, but the repeating module has a different structure. Instead of a simple neuron, it is a cell. The information flow of A LSTM cell is regulated by structures called gates as illustrated in figure \ref{fig:pic_lstm}.
\hfill\break

The LSTM is trained to accept a sequence of the corrected hit positions (X, Y, and Z), and outputs a probability that the input sequence belongs to the positive class (true track) or not. The best results of validation accuracy was obtained by the stacking of 3 LSTM layers one after another. In addition to that, we applied a 50\% dropout layer to each layer to prevent our model from over-fitting. After 1000 training iterations our LSTM network scores 95\% accuracy.

\label{subsection}

\subsection{Performance:}
\begin{figure}[!htb]
  \centering
  \includegraphics[width=0.85\linewidth]{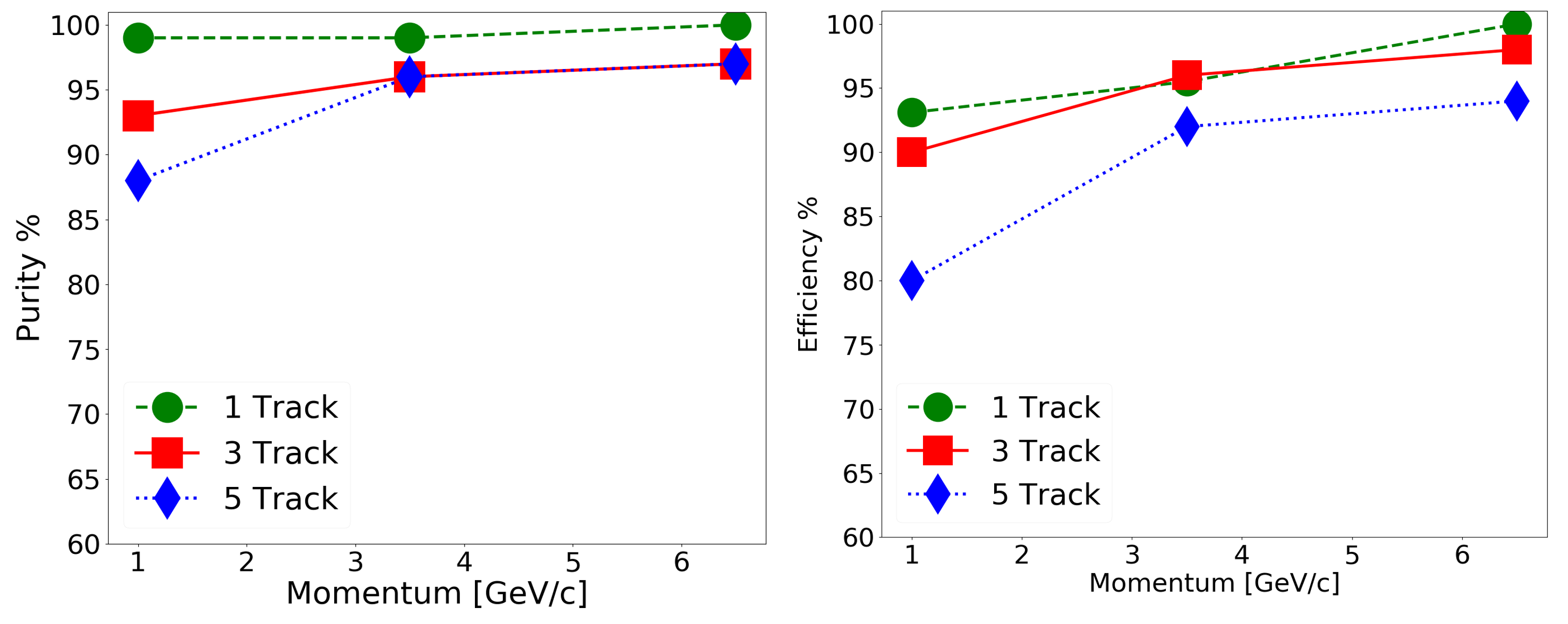}
  \caption{Purity and Efficiency of the whole algorithm. }
  \label{fig:pic_eff}
\end{figure}

The purity of the proposed algorithm is defined as the fraction of hits in one track come from the correct particle. The efficiency is defined as the ratio of the number of simulated tracks with more than 30 hits for low momentum (\textless 1.5 GeV/c) tracks passing through (FT1,FT2), (FT3,FT4), and more than 38 hits for tracks with higher momenta, passing through all the stations, to the number of all simulated tracks. Figure \ref{fig:pic_eff} shows the purity and efficiency of the algorithm as a function of momentum for the three different track multiplicities. For high momentum tracks (6.5 GeV/c), purity and efficiency is close to 100\%, for moderate momentum tracks (3.5 GeV/c) the purity is above 95\%, while the efficiency is above 90\%. For low momentum tracks (1 GeV/c) efficiency drops to 80\% for the 5 track multiplicity. 

\label{subsection}
 
\section{Conclusion}

The goal of the work presented in this paper is the design of a pattern recognition algorithm for the PANDA FTS detector. The current status
is presented in form of a complete algorithm alongside a description of a proof-of-concept implementation. The milestones for further developments foresee the finalization and test of a full algorithm implementation. We expect, that the efficiency can still be improved, since our algorithm has been finished very recently and contains several parameters and conditions that can be further optimized.


\Acknowledgements
The authors acknowledge support from the J{\"u}lich Research Center. W. Esmail would like to thank Micheal Kunkel and Daniel Lersch for the fruitful discussions. 




\newpage
\begin{thebibliography}{99}


\bibitem{panda} 
  C. Schwarz for the PANDA Collaboration,
  "The PANDA Experiment at FAIR",
  J. Phys. Conf. Ser. {\bf 374}, Pages 012003 (2012)
  
\bibitem{physics} 
  U. Wiedner for the PANDA Collaboration,
  "Future prospects for hadronp Physics at PANDA",
  Pro. Part. Nucl. Phys. {\bf 66}, Pages 477-518 (2011)

\bibitem{fts} 
  PANDA Collaboration,
  "Technical Design Report for the PANDA Forward Tracker",
  (2018), URL: \url{https://panda.gsi.de/system/files/user_uploads/admin/RE-TDR-2017-001.pdf}
  
\bibitem{straw} 
  Sedigheh Jowzaee,
  PhD Thesis "Self-Supporting Straw Tube Detectors for the COSY-TOF and PANDA Experiments",
  (2014), URL: \url{http://juser.fz-juelich.de/record/185801/files/published-Jowzaee-PhD-dissertation-UJ-Physics.pdf}
  
\bibitem{pandaroot} 
  S. Spataro for the PANDA Collaboration,
  "The PandaRoot framework for simulation, reconstruction and analysis",
  J. Phys. Conf. Ser. {\bf 331}, Pages 032031 (2011)
  
\bibitem{keras} 
  Chollet, Fran\c{c}ois and others,
  "Keras",
  (2015), URL: \url{https://keras.io}
  
\bibitem{entropy} 
  S. Kullback, R. Leibler,
  "On information and sufficiency",
  Ann. Math. Stat. {\bf 22}, Pages 79-86 (1951)
  
\bibitem{dropout} 
  N. Srivastava et al.,
  "Dropout: a simple way to prevent neural networks from overfitting",
  J. Mach. Learn. Res. {\bf 15}, Pages 1929-1958 (2014)
  
\bibitem{odr} 
  Boggs et al.,
  "User's Reference Guide for ODRPACK Version 2.01 Software for Weighted Orthogonal Distance Regression.",
  NISTIR 4834 (1992), URL: \url{https://docs.scipy.org/doc/external/odrpack_guide.pdf}
  
\bibitem{scipy} 
  Eric Jones et al.,
  "{SciPy}: Open source scientific tools for {Python}",
  (2001), URL: \url{http://www.scipy.org/}
  
\bibitem{lstm} 
  S. Hochreiter and J. Schmidhuber,
  "Long short-term memory",
  Neural Comput. {\bf 9(8)}, Pages 1735–1780 (1997)
  
\bibitem{lstmcell} 
  Christopher Olah,
  "Understanding LSTM Networks",
  URL: \url{https://colah.github.io/posts/2015-08-Understanding-LSTMs/}   

\end{thebibliography}
\end{document}